\documentclass{article}
\usepackage{amsfonts}
\usepackage{amssymb}
\usepackage{graphicx}
\usepackage{amsmath}

\input{tcilatex}

\begin{document}

\title{ {\bf Minimal duality breaking in the Kallen-Lehman approach to 3D Ising
model: a numerical test}}
\author{Marco Astorino$^{1,4}$ \thanks{%
marco.astorino AT gmail.com} , Fabrizio Canfora$^{1}$ \thanks{%
canfora AT cecs.cl}, \and Cristi\'an Mart\'{\i}nez$^{1,5}$ \thanks{%
martinez AT cecs.cl} , Luca Parisi$^{2,3}$ \thanks{%
parisi AT sa.infn.it} \\
\\
{\small $^{1}$\textit{Centro de Estudios Cient\'{\i}ficos} (\textit{CECS),
Valdivia, Chile}}\\
{\small $^{2}$\textit{Istituto Nazionale di Fisica Nucleare, GC di Salerno,
Italy}}\\
{\small $^{3}$\textit{Dipartimento di Fisica E. R. Caianiello, Universit\`{a}
di Salerno, Italy}}\\
{\small $^{4}$\textit{Instituto de F\'{\i}sica, Pontificia Universidad
Cat\'olica de Valpara\'{\i}so, Chile}}\\
{\small $^{5}$\textit{Centro de Ingenier\'{\i}a de la Innovaci\'{o}n del
CECS (CIN), Valdivia, Chile} }}
\maketitle

\begin{abstract}
A Kallen-Lehman approach to 3D Ising model is analyzed numerically both at
low and high temperature. It is shown that, even assuming a minimal duality
breaking, one can fix three parameters of the model to get a very good
agreement with the MonteCarlo results at high temperatures. With the same
parameters the agreement is satisfactory both at low and near critical
temperatures. How to improve the agreement with MonteCarlo results by
introducing a more general duality breaking is shortly discussed.



CECS-PHY-08/01
\end{abstract}

\section*{Introduction}

The three dimensional Ising model (henceforth 3DI) is one of the main open
problems in field theory and statistical mechanics. A large number of
interesting statistical systems near the transition point are described by
3DI and the theoretical methods suitable to deal with such a problem
manifest deep connections in various areas of physics ranging from quantum
information theory to string theory (for a review see e.g. \cite{PV02}).
Besides its intrinsic interest in statistical physics, since the formulation
of the Svetitsky-Yaffe conjecture \cite{SV82}, it has been widely recognized
its role in describing the deconfinement transition in QCD. For this reason,
the 3D Ising model is worth to be further investigated.

It has been recently proposed \cite{CPV06} to use in statistical mechanics
the powerful tools of Regge theory \cite{Re1, Re2} which have been so
fruitful in the study of the strong interactions leading to the formulation
of the dual models \cite{Ve1} (two detailed reviews are \cite{Co71} and \cite%
{Ve74}). It has been argued in \cite{Ca07} that such ideas may be useful in
dealing with the 3D Ising model. It would be also interesting to try to test
these methods in the mean field theory of spin glass\footnote{%
Detailed reviews on the subject are \cite{MPV87} \cite{Ni02} \cite{CC05}.}
as developed in \cite{Pa79} (recently\ it has been proved, that such a
framework provides one with the exact solution in the mean field Ising spin
glass \cite{Gu02, GT02, Ta03}).

In the present paper we perform a numerical test of these methods in the
case of the 3D Ising model. We will try to find the best "Regge parameters"
appearing in the Kallen-Lehman form for the free energy: that is the
parameters which provide one with the best possible agreement both at high
and low temperatures MonteCarlo expansions as well as around the critical
point.

The paper is organized as follows: in the second section the Kallen-Lehman
form for the free energy of the 3D Ising model is discussed. In the third
section the issue of duality breaking is analyzed. In the fourth and fifth
sections the comparison with MonteCarlo results at high and low temperatures
are presented. In the sixth section, the behavior at the critical point is
described. In the seventh section some possible improvements are pointed
out. Eventually, we draw some conclusions.

\section{The Kallen-Lehman form for the free energy}

The Kallen-Lehman representation \cite{Ca07} gives rise to an ansatz for the
free energy of the 3DI model of the following form\footnote{%
The "Regge parameters" appearing in these formulas are related with those
appearing in \cite{Ca07} by the following identities $\zeta _{1}=\nu $, $%
\zeta _{3}=1/2=\zeta _{2}$, $\zeta _{0}=\alpha $.} 
\begin{eqnarray}
F_{3D}^{(\zeta _{i},\lambda )}(\beta ) &=&F_{2D}(\beta )+\frac{\lambda }{%
\left( 2\pi \right) ^{2}}\int\limits_{0}^{\pi }dz\int\limits_{0}^{\pi
}dy\cdot  \notag \\
&&\cdot \log \left\{ \frac{1}{2}\left[ 1+\left( 1-\left[ 2\frac{\left(
\Delta (z)-1\right) ^{\zeta _{1}}}{\Delta (z)}\right] ^{\zeta _{2}}\sin
^{2}y\right) ^{\zeta _{3}}\right] \right\} ,  \label{parto32} \\
F_{2D}(\beta ) &=&F_{1D}(\beta )+\frac{1}{2\pi }\int\limits_{0}^{\pi }dx\log
\left\{ \frac{1}{2}\left[ 1+\sqrt{1-k_{eff}(\beta )^{2}\sin ^{2}x}\right]
\right\} ,  \label{t2D} \\
F_{1D}(\beta ) &=&\log 2\cosh q\beta ,\ \ \Delta (z)=\left( 1+\left(
1-k_{eff}(\beta )^{2}\sin ^{2}z\right) ^{\zeta _{0}}\right) ^{2},\ 
\label{trivial} \\
\zeta _{1},\zeta _{2},\zeta _{3} &>&0,\ \ \ \ q=2  \notag
\end{eqnarray}%
here, for sake of simplicity, we consider $\zeta _{1}=1/2=\zeta _{3}$ (the
2D Ising values). In order to avoid a complex free energy (which could arise
due to the two square roots), 
\begin{eqnarray}
0 &\leq &k_{eff}(\beta )\leq 1\Rightarrow  \label{condPT} \\
\ \ \quad 1 &\leq &\Delta (z)\leq 4\ \ \vee \ \ \zeta _{2}>0\Rightarrow 
\notag \\
0 &\leq &\left[ 2\frac{\left( \Delta (z)-1\right) ^{1/2}}{\Delta (z)}\right]
^{\zeta _{2}}\leq 1,  \label{lowte11} \\
u &=&\exp (-2\beta ),\ \ t=\tanh \beta .  \label{lowte2}
\end{eqnarray}%
At high temperatures it is convenient the variable $t=\tanh \beta $ while $%
u=\exp (-2\beta )$ is the standard variable at small temperatures. Thus our
trial free energy will depend on 3 parameters $\lambda ,\ \zeta _{2}:=\nu ,\
\zeta _{0}:=\zeta $ to be fixed in comparison with MonteCarlo data.

The natural value (suggested by the application of Regge theory) of the
factor $q$ in the trivial one dimensional term in Eq. (\ref{trivial}) is $%
q=2 $ (and this is the value that we will consider from now on, see \cite%
{Ca07}) since the one dimensional term of exact solution of the 2D Ising
model is precisely $\log 2\cosh 2\beta $. On the other hand, we have also
obtained quite good numerical results for $q=1$: in particular, for the set
of parameters 
\begin{equation}
\zeta _{q=1}^{\ast }=1.096,\ \ \nu _{q=1}^{\ast }=2.586,\ \ \lambda
_{q=1}^{\ast }=0.127  \label{old}
\end{equation}%
we get the following deviations at high and small temperatures%
\begin{eqnarray}
\sigma _{HT}(\zeta _{q=1}^{\ast },\nu _{q=1}^{\ast },\lambda _{q=1}^{\ast })
&\approx &\sqrt{\frac{\chi (\zeta _{q=1}^{\ast },\nu _{q=1}^{\ast },\lambda
_{q=1}^{\ast })}{50}}\approx 10^{-6}  \label{deh0} \\
\sigma _{LT}(\zeta _{q=1}^{\ast },\nu _{q=1}^{\ast },\lambda _{q=1}^{\ast })
&\approx &\frac{\sqrt{\chi _{LT}(\zeta _{q=1}^{\ast },\nu _{q=1}^{\ast
},\lambda _{q=1}^{\ast })}}{\sqrt{50}}\approx 10^{-3}  \label{del0}
\end{eqnarray}%
(a definition of $\sigma _{HT}$ and $\sigma _{LT}$ will be provided in the
next sections in Eqs. (\ref{errhigh0}), (\ref{errhigh}), (\ref{comme2}) and (%
\ref{errlow})). These quite small deviations obtained with $q=1$ (which
indeed is less natural than the ansatz with $q=2$) show that the present
method is well suited to describe the thermodynamics of the 3D Ising model.
In any case, as it will be shown in a moment, the results for $q=2$ are
quite better.

Besides the first trivial term in Eq. (\ref{trivial}), the "Regge free
energy" depends on the temperature through an effective $k_{eff}(\beta )$ in
the same way as the (non trivial part of the) free energy of the 2D Ising
model depends on $\beta $: 
\begin{equation}
\left( k_{2D}(\beta )\right) ^{2}=\left( \frac{2}{\cosh 2\beta \coth 2\beta }%
\right) ^{2}=\left( 4\frac{\exp (2\beta )-\exp (-2\beta )}{\left( \exp
(2\beta )+\exp (-2\beta )\right) ^{2}}\right) ^{2}.\quad  \label{ka2}
\end{equation}%
The duality symmetry of the 2D Ising model (discovered in \cite{KW41} by
Kramers and Wannier before the exact solution of Onsager \cite{O44}) is
manifest in the above function of the temperature since $k(\beta )$ has the
same expression when rephrased in terms of the low temperatures variable $u$
and the high temperatures variable $t$. Indeed, the 3D Ising model has not
such a duality symmetry: in the next section it will be described how the
simplest duality breaking can be achieved.

\section{Minimal duality breaking}

There are indeed many reasonable ways to break duality, how can one choose?
Here a criterion of simplicity will be followed. An important ingredient is
the following: all the Ising models on simple hypercubic lattices are
invariant under the Marchesini-Shrock transformation \cite{MS88}%
\begin{equation*}
\beta \rightarrow \beta +in \frac{\pi}{2} ,\ \ n\in 
\mathbb{Z}
.
\end{equation*}%
While this symmetry is obviously realized in the 2D case in which one has at
own disposal the exact solution, such an exact symmetry puts a strong
constraint on the form of the free energy. As discussed in \cite{Ca07}, the
free energy suggested by the Kallen-Lehman representation is expect to
depend on (a suitable) $k_{eff}(\beta )$. A simplicity criterion (which, as
it will be shown in a moment, is supported by the numerical data) suggests
to consider the easiest possible modification of the $k$ of the 2D case in
Eq. (\ref{ka2}):%
\begin{equation}
k_{eff}(\beta )=\frac{4}{d_{1}}\frac{d_{3}\exp (2\beta )-d_{2}\exp (-2\beta )%
}{\left( \exp (2\beta )+d_{0}\exp (-2\beta )\right) ^{2}},  \label{kaeff}
\end{equation}%
the fact that the exponential only depend on $2\beta $ is the simplest
possible way to fulfil the Marchesini-Shrock symmetry. The four parameters $%
d_{i}$ ($i=0,.., 3$) are not independent since there are two conditions that
the function $k_{eff}$\ has to fulfil: firstly, the maximum of $k_{eff}$ has
to occur at the known critical temperature $\left( \beta ^{\ast }\right)
^{-1}$ of the 3D Ising model:%
\begin{eqnarray}
\left. \beta ^{\ast }\right\vert \ \left. \partial _{\beta }k_{eff}(\beta
)\right\vert _{\beta =\beta ^{\ast }} &=&0  \label{co11} \\
\beta ^{\ast } &=&0.22165.  \label{co12}
\end{eqnarray}%
The second condition is related to the fact that the expected transition has
to occur \cite{Ca07} when $k_{eff}(\beta )=1$ as it happens in the 2D case:%
\begin{equation}
k_{eff}(\beta ^{\ast })=1.  \label{co2}
\end{equation}%
Thus, in order to consider the simplest modification of the two-dimensional $%
k(\beta )$ in Eq. (\ref{ka2}), we will take $d_{3}=1$ (the 2D Ising value)
in such a way that only one parameter is left: $d_{1}$ and $d_{2}$ can be
expressed in terms of $d_{0}$ and $\beta ^{\ast }$ as follows%
\begin{eqnarray}
d_{2} &=&\frac{\exp (4\beta ^{\ast })-3d_{0}}{3-d_{0}\exp (-4\beta ^{\ast })}%
;  \label{co3} \\
d_{1} &=&4\left[ \frac{\exp (2\beta ^{\ast })-\left( \frac{\exp (4\beta
^{\ast })-3d_{0}}{3-d_{0}\exp (-4\beta ^{\ast })}\right) \exp (-2\beta
^{\ast })}{\left( \exp (2\beta ^{\ast })+d_{0}\exp (-2\beta ^{\ast })\right)
^{2}}\right] .  \label{co4}
\end{eqnarray}%
From the numerical point of view, this parametrization is very useful since
automatically ensures that the maximum is equal to one avoiding possible
problems related to negative numbers appearing inside the square root in Eq.
(\ref{t2D}). Moreover the minimal modification criterion suggests that 
\begin{equation}
d_{0}=1  \label{kasimple}
\end{equation}%
as in the 2D case.

Indeed, one should expect a more general duality breaking: the 2D Ising
model and its exact duality are closely related to \textbf{N}=4 \ SUSY
Yang-Mills theory (for instance see \cite{GGH07}). From the gauge theory
side, one would thus expect that a minimal duality breaking should be
related to a Yang-Mills theory with two supersymmetry which has an effective
duality (for instance see \cite{OW99}). The 3D Ising model is related to QCD
without supersymmetries, so that it is natural to expect that a more general
pattern of duality breaking (which will be discussed later on) should occur
in the 3D Ising model. However, it is a truly remarkable feature of the
present tools that already a minimal duality breaking works very well in
comparisons with MonteCarlo results. It is worth to mention here that the so
called "effective string approch" to the 3D Ising model (in which duality
symmetry is unbroken) gives reasonable resuts in comparison with MonteCarlo
data (see for instance \cite{pan03}).

One can express $k_{eff}$ in terms of the high and low temperatures
variables:%
\begin{eqnarray}
k_{eff}(u) &=&\frac{4}{d_{1}}\frac{u\left( 1-d_{2}u^{2}\right) }{\left(
1+d_{0}u^{2}\right) ^{2}},  \label{kalow} \\
k_{eff}(t) &=&\frac{4\left( 1-t^{2}\right) }{d_{1}}\frac{\left[ t^{2}\left(
1-d_{2}\right) -2t\left( 1+d_{2}\right) +\left( 1-d_{2}\right) \right] }{%
\left[ t^{2}\left( 1+d_{0}\right) -2t\left( 1-d_{0}\right) +\left(
1+d_{0}\right) \right] ^{2}}.  \label{kahigh}
\end{eqnarray}%
so our ansatz for the 3D Ising free energy is given by Eq. (\ref{parto32})
in which $k_{eff}$\ is in Eq. (\ref{kaeff}) with the constants in Eqs. (\ref%
{co3}), (\ref{co4}) and (\ref{kasimple}).

\section{High temperatures}

The idea is to find the best set of high temperatures parameters $\left(
\zeta _{0}^{\ast },\zeta _{2}^{\ast },\lambda ^{\ast }\right) $ in Eqs. (\ref%
{parto32}), (\ref{trivial}) (with the constraints in Eqs. (\ref{co3}), (\ref%
{co4}) and (\ref{kasimple})) which reproduces as close as possible the
available Monte Carlo data (see \cite{AF02}). A hypercubic lattice has been
chosen in the parameters space (every point in the lattice representing a
possible set of high temperature parameters), then the free energy (\ref%
{parto32}) will be evaluated at every point in the lattice. The best choice
of parameters will be the one minimizing the following deviation function
which, to some extent, represents the deviation between the ansatz and the
Monte Carlo data: 
\begin{equation}
\chi (\zeta ,\nu ,\lambda )=\sum\limits_{i}^{50}\left\vert F_{3D}^{(\zeta
,\nu ,\lambda )}(\beta _{i})+I_{0}-\left. F\right. _{HT}^{\ \ MC}(\beta
_{i})\right\vert ^{2},\   \label{errhigh0}
\end{equation}%
\begin{equation*}
\beta _{i}-\beta _{i-1}=\frac{0.03}{50}\ \ ,\qquad \beta _{50}=\beta
_{max}=0.03,\ \ \ I_{0}=2.4831
\end{equation*}%
\begin{eqnarray*}
\left. F\right. _{HT}^{\ \ MC}(\beta ) &=&3\cosh \beta +\left( 3\left( \tanh
\beta \right) ^{4}+22\left( \tanh \beta \right) ^{6}+187.5\left( \tanh \beta
\right) ^{8}+\right. \\
&&\left. 1980\left( \tanh \beta \right) ^{10}+24044\left( \tanh \beta
\right) ^{12}+319170\left( \tanh \beta \right) ^{14}+...\right)
\end{eqnarray*}%
where we keep the terms up to the $15th$ of \cite{AF02} since our algoritm
is not sensitive to the higher order terms, $\left. F\right. _{HT}^{\ \ MC}$
is the high temperatures MonteCarlo free energy, $\beta _{m}$ can be assumed
to be of order $0.03$\footnote{%
After $\beta \approx 0.05$ one is not anymore in the high temperatures
regime since the critical temperature is at $\beta ^{\ast }\approx 0.22$
which is only a factor of four larger. Indeed, $\beta \approx 0.03$ appears
to be not enough smaller than $\beta ^{\ast }$. Nevertheless, we will see
that the agreement of the Regge free energy with MonteCarlo data is
excellent up to $\beta \approx 0.03$.} and $I_{0}$ is a constant introduced
for numerical convenience\footnote{%
In order for the program to be able to select good points in the parameters
space an arbitrary constant has to be added. Otherwise, the $C++$ program
would select uncorrect functions: for instance, it would select two curves
which intersect in a point near $\beta =0$ (but whose shapes are very
different) instead of two parallel curves.}.

In order to compare the Regge coefficients with the high temperature
coefficients in \cite{AF02}, we need to change variable from $\beta $ to $%
t=\tanh \beta $, so that we have to use the expression in Eq. (\ref{kahigh})
whenever it appears $k_{eff}$.

The set of points candidates to be the best parameters are: 
\begin{equation}
\zeta ^{\ast }=1.9389,\ \ \nu ^{\ast }=1.9205,\ \ \frac{\lambda ^{\ast }  }{ (2\pi)^2 } = 0.0781  
\label{OHT}
\end{equation}

In order to have a graphical idea of the closeness of the Regge and the
MonteCarlo free energies at high temperature we have drawn in Fig. (\ref%
{fig-ht}) both functions. The agreement appears to be excellent: the
deviation at high temperature is 
\begin{equation}
\sigma _{HT}(\zeta ^{\ast },\nu ^{\ast },\lambda ^{\ast })\approx \sqrt{%
\frac{\chi (\zeta ^{\ast },\nu ^{\ast },\lambda ^{\ast })}{50}}\approx
10^{-6}.  \label{errhigh}
\end{equation}

It is worth to note that in the range of parameters%
\begin{equation*}
1\leq \zeta ^{\ast }\leq 2,\ 1.5\leq \ \nu ^{\ast }\leq 2.5,\ \ 0.05\leq
\lambda ^{\ast }\leq 0.2
\end{equation*}%
we found many other good sets of parameters with only slightly larger
deviations with respect to the MonteCarlo free energy. However, as it will
be discussed in a moment, it is very difficult to evaluate the errors on the
single coefficients of the expansions. Thus, we simply considered the set in
Eq. (\ref{OHT}) but other sets of parameters in the above range could
reproduce better the single coefficients.

\begin{figure}[!htp]
\caption{MonteCarlo $F_{HT} ^{\ \ \ MC}(\protect\beta )$ and Kallen-Lehman $%
F_{3D}^{(\protect\zeta ^{\ast },\protect\nu ^{\ast },\protect\lambda ^{\ast
})}(\protect\beta)$ free energies at high temperatures versus $\protect\beta$%
.}
\label{fig-ht}
\begin{center}
\includegraphics[angle=270, scale=0.4]{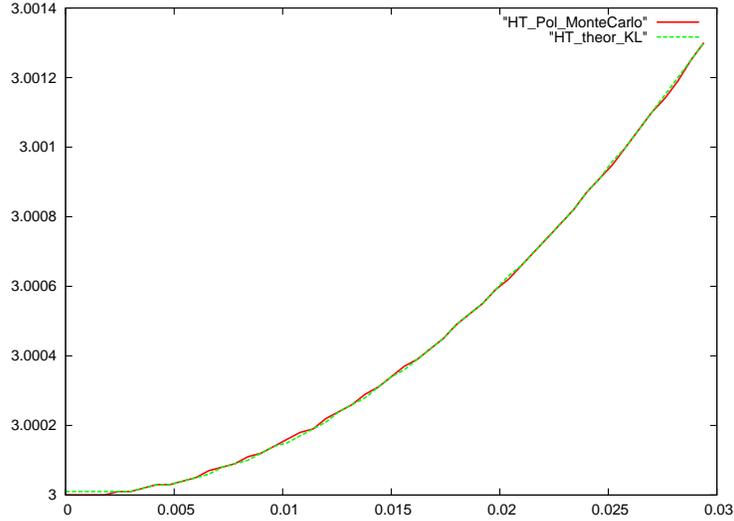}
\end{center}
\end{figure}

There are some points in the parameters space in which the agreement at low
temperatures is much better than the agreement at high temperatures (which
is in any case quite good). In the present analysis it is apparent that one
cannot achieve an excellent agreement both at high and at low temperatures:
in the best cases very small deviations (of the order of $10^{-6}$) on one
side correspond on the other side to quite good but not excellent deviation
of the order of $10^{-4}$. This is a manifestation of the already discussed
fact that the 3D Ising model, being related to QCD without any
supersymmetry, needs a more general duality breaking.

\subsection{Numerical Derivatives at high temperatures}

A consistency test would be provided comparing the derivatives of MonteCarlo
free energy with the ones predicted by our model at $\beta=0$. Due to the
complicated expression of the Regge free energy numerical methods are
needed. To this aim we consider the finite difference approximation of the
derivative which, at any order of derivation $n$, is given by the forward
finite difference of order $n$ divided by the increase of the independent
variable (say $\delta $) raised to the $n$-th power, i.e. 
\begin{equation*}
\frac{d^{n}F_{3D}^{(\zeta ^{\ast },\nu ^{\ast },\lambda ^{\ast })}}{d\beta
^{n}}(\beta )=\frac{\Delta _{\delta }^{n}[F_{3D}^{(\zeta ^{\ast },\nu ^{\ast
},\lambda ^{\ast })}](\beta )}{\delta ^{n}}+O(\delta )
\end{equation*}%
$\Delta _{\delta }^{n}[f](x)$ being the $n$-th order finite difference.
Behind this apparently easy procedure there are hidden dangers which already
at the lowest orders deserve special attention (for an introductory
discussion see \cite{NR07}): which is the best choice of $\delta $ in order
to approximate correctly the value of the $n$-th order derivative (since
neither too small nor too large values are correct)? One can only try to
find the suitable $\delta _{(n)}$ to reproduce the MonteCarlo coefficients
at any order. In the case of the Regge free energy it's always possible to
find (more than) one suitable numerical value for $\delta $. One can verify
that the $\delta _{(n)}$ which reproduce the MonteCarlo results fulfil a
sort of "natural" scaling (see \cite{NR07}): namely, as the order of
derivation increases, the suitable $\delta $ reproducing the MonteCarlo
result stays constant or decreases. Despite being only a first attempt to
compute the derivatives of the Regge partition function in zero,
nevertheless it is a quite encouraging fact that at any order one can find
at least one $\delta $ reproducing exactly the MonteCarlo data.

Another possible way to obtain the numerical derivatives for $\beta =0$ is
the following. One can fit the Regge free energy (in which now the
parameters are fixed to be the best parameters) for small $\beta $ with a
polynomial with unknown coefficients (in which \textit{a priori} one has to
include all the-even and odd-coefficients) in such a way that the
coefficients of the polynomial are the numerical derivatives at the origin.
However, after the first four coefficients are fixed, the polynomial becomes
very close to the Regge free energy. The higher order coefficients are fixed
by the program to be the MonteCarlo results, but it is very difficult to
evaluate the errors since if one changes the higher order coefficients of a
factor of two or four the deviation with respect to the optimal polynomial
is smaller than $10^{-6}$. Even with this method, it is extremely difficult
to evaluate the deviations of the derivatives of the higher order
derivatives of the Regge free energy from the MonteCarlo results. In any
case, we find an attractive feature of our model that even fixing the first
few coefficients at high temperature one obtains a good agreement at the
critical point and at low temperatures.

\section{Low temperatures}

Once the best set of high temperatures parameters $\left( \zeta ^{\ast },\nu
^{\ast },\lambda ^{\ast }\right) $ have been found in Eq. (\ref{OHT}) one
could hope to find a good agreement also at low temperatures (something
which, \textit{a priori}, is very far from obvious). The internal energy at
low temperatures is ($k_{eff}$ as in Eq. (\ref{kalow}) and expressing $%
F_{1D} $ as in (\ref{trivial}) in terms of $u$) 
\begin{equation}
\left\langle \frac{E}{N}\right\rangle ^{KL}(u)+ 2 I_{1}=2u\frac{\partial }{%
\partial u}F_{3D}^{(\zeta ,\nu ,\lambda )}(u) \ .  \label{parto33}
\end{equation}
This is the average energy for spin (the two expressions can differ by a non
zero constant $I_{1}$, see footnote 4) to be compared with $\left\langle 
\frac{E}{N}\right\rangle ^{MC}$ (the polynomial in $u$ representing the
MonteCarlo average energy for spin for small $u$) found in \cite{BCL92}.

It is worth to discuss the origin of the constant $I_{1}$ in the above
formula. The partition function of the 3D Ising model reads 
\begin{eqnarray*}
Z_{3D} &=&\sum_{\left\{ \sigma _{i}\right\} }\exp \left( -\beta \left(
H_0\right) \right) , \\
H_0 &=&\sum_{\left\langle ij\right\rangle }\sigma _{i}\sigma _{j}+const
\end{eqnarray*}%
where the constant in the Hamiltonian can always be chosen in such a way
that the lowest energy state has zero energy. The ansatz (\ref{parto32}) has
not been deduced by solving in some approximated way the model. Rather, it
has been found on the basis of the Regge theory and the Kallen-Lehman
representation: one cannot pretend that it already corresponds to the
normalization in which the Hamiltonian $H_{0}$ has a zero energy ground
state. At very low temperatures, the free energies corresponding to $H_{0}$
and to $H_{0}+const$ differ by a term proportional to $1/T$%
\begin{equation*}
\log \left[ \sum_{\left\{ \sigma _{i}\right\} }\exp \left( -\beta \left(
H_{0}+const\right) \right) \right] -\log \left[ \sum_{\left\{ \sigma
_{i}\right\} }\exp \left( -\beta \left( H_{0}\right) \right) \right] \sim
\beta \sim 1/T\sim \log u.
\end{equation*}%
In terms of the internal energy, such a term corresponds to a constant:%
\begin{equation*}
2u\frac{\partial }{\partial u}\log u=const.
\end{equation*}

From the numerical point of view it is extremely inconvenient to use the
expression on the right hand side of Eq. (\ref{parto33}) since the
derivative $\partial _{u}$ has a really big expression. A trivial but useful
trick is the following: in \cite{BCL92} one has a polynomial expression for
the average energy for spin so that one can integrate it and obtain the
Monte Carlo expression for the free energy at low temperatures:%
\begin{equation*}
\left\langle \frac{E}{N}\right\rangle ^{KL}(u)+ 2 I_{1}=2u\frac{\partial }{%
\partial u}F_{LT}^{\ \ \ MC}(u)=\sum_{i=6}^{14}a_{i}^{(L)}u^{i},
\end{equation*}%
this implies that%
\begin{equation}
F_{LT}^{\ \ \ MC}(u)=\frac{1}{2}\left( \sum_{i=6}^{14}\left( \frac{a_{i}^{(L)}}{i}%
\right) u^{i}- 2 I_{1}\log u\right) .  \label{lowFE}
\end{equation}%
Therefore, the low temperatures test function reads%
\begin{equation}
\chi _{LT}(\zeta ^{\ast },\nu ^{\ast },\lambda ^{\ast
})=\sum_{i=1}^{50}\left\vert F_{3D}^{(\zeta ^{\ast },\nu ^{\ast },\lambda
^{\ast })}(u_{i})+I_{2}-F_{LT}^{\ \ \ MC}(u_{i})\right\vert ^{2},
\label{comme2}
\end{equation}%
\begin{equation*}
u_{i}-u_{i-1}=\frac{0.3}{50},\ \ \ \ u_{1}=u_{min},\ \ \ \ u_{50}=u_{\max
}=0.3,
\end{equation*}%
\begin{eqnarray*}
F_{LT}^{\ \ \ MC}(u) &=&\left[ \frac{1}{2}\left( \frac{12}{6}\left( u\right)
^{6}+\frac{60}{10}\left( u\right) ^{10}+\right. \right. \\
&&\left. \left. -\frac{84}{12}\left( u\right) ^{12}+\frac{420}{14}\left(
u\right) ^{14}...\right) +I_{1}\log u\right]
\end{eqnarray*}%
where $I_{1}=-1$ and $I_{2}=0.1034$. We keep the terms up to the $15th$ of 
\cite{BCL92}\ since our algoritm is not sensitive to the higher order terms
and $u_{\max }$ can be assumed to be of order $0.3$.

Actually, the value of $u$ which corresponds to the critical temperature is $%
u_{crit}=\exp (-2\beta _{crit})\approx 0.6$. The low temperatures expansion
provides one with reasonable results for $u\ll 0.6$. Indeed, $u_{\max
}\approx 0.3$ is not at all much smaller than $u_{crit}$: one may expect a
reasonable value of $u_{\max }$ to be quite smaller than\footnote{%
Unfortunately, a numerical analysis in the region of $u<0.01$ is quite
difficult because of the fact that the logaritm is very large and dominates
the too small numbers coming from the MonteCarlo polynomial. If the
logartmic divergent terms would be absent, in the region $0\leq u<0.01$ very
soon the MonteCarlo polynomial would give rise to very small numbers (below
the precision of our PC). Thus, to analyze this range more refined numerical
techniques are needed. In any case, the good agreement at the critical point
provides one with some control on the deviation of the theoretical low
temperatures coefficients with respect to the MonteCarlo ones.} $0.01$.
Nevertheless, we will see that the agreement is quite good up to $u\approx
0.3$.

Eventually, the deviation at low temperatures between the Regge and the
MonteCarlo free energies \textit{evaluated for the same optimal parameters}
in Eq. (\ref{OHT}), which have been found by asking the optimal agreement at
high temperatures is%
\begin{equation}
\sigma _{LT}(\zeta ^{\ast },\nu ^{\ast },\lambda ^{\ast })\approx \frac{%
\sqrt{\chi _{LT}(\zeta ^{\ast },\nu ^{\ast },\lambda ^{\ast })}}{\sqrt{50}}%
\approx 10^{-5}.  \label{errlow}
\end{equation}%
and one can see that the agreement at low temperatures is two order of
magnitude better than in the case in which in Eq. (\ref{trivial}) $q=1$ (see
Eq. (\ref{del0})).

\begin{figure}[!htp]
\caption{MonteCarlo $F_{LT}^{\ \ \ MC}$  and Kallen-Lehman $F_{3D}^{(\protect\zeta ^{\ast },%
\protect\nu ^{\ast },\protect\lambda ^{\ast })}(u)$ free energies at low
temperatures versus $u$.}
\label{fig-lt}
\begin{center}
\includegraphics[angle=270, scale=0.4]{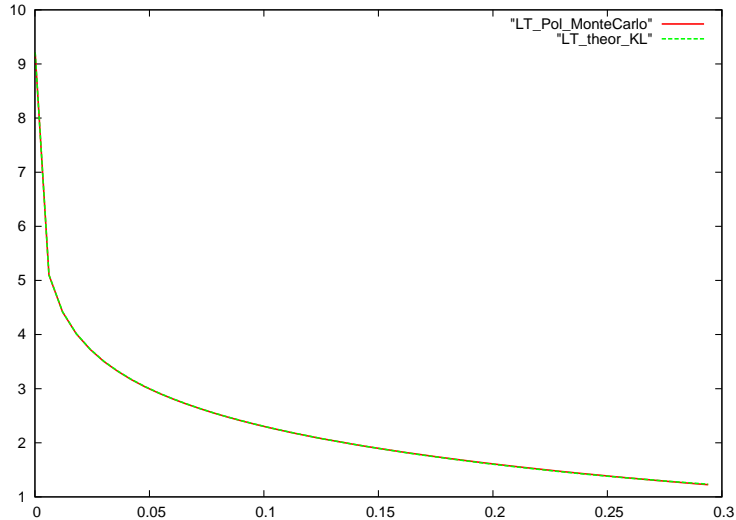}
\end{center}
\end{figure}

Remarkably enough, at low temperatures also the agreement is quite good as
can be also seen in Fig. (\ref{fig-lt}). As far as the numerical derivatives
at $u=0$ of Regge free energy, the same considerations as in the high
temperatures case hold. Namely, both a direct evaluation by means of a
simple algoritm and to try to find the suitable approximating polynomial
give results compatible with the MonteCarlo ones but it is not clear to the
present authors how to estimate the errors on the single coefficients. It is
nevertheless worth to stress here that the relative deviations both at high
and at low temperatures are very small (see Eqs. (\ref{errhigh}) and (\ref%
{errlow})). Furthermore, as it will be now shown, also the critical exponent
is in a very good agreement with the available data: one may hope that the
good agreement at the critical point could prevent too large deviations of
the the Kallen-Lehman derivatives from the MonteCarlo data.

\section{The Critical Point}

The last test is the critical point. Namely, once the parameters have been
fixed as in Eq. (\ref{OHT}) one may hope to verify that the critical point
is correctly predicted too. To do this one can fit near the critical point
the non-analytic part of the free energy in Eq. (\ref{parto32})\footnote{%
That is, one has to exclude the term $\log 2\cosh \beta $ which does not
contribute to the critical behavior.} with the optimal parameters with a
function of the form $F_{CRIT}$%
\begin{equation}
F_{CRIT}\approx c_{1}\left\vert \beta -\beta ^{\ast }\right\vert ^{2-\alpha
}+c_{2}  \label{critest}
\end{equation}%
(where $c_{1}$, $c_{2}$ and $\alpha $ are constant) and find the optimal
values for $c_{1}$, $c_{2}$ and $\alpha $ so that $\alpha $ will be our
estimate for the critical exponent. This form of the free energy's critical
part is expected both from Conformal Field Theory and experiments. The
results of the fit done with the PC program \textit{Mathematica} by fitting
the (non-analytic part of the) Regge free energy with the optimal parameters
in Eq. (\ref{OHT}) with the above function (\ref{critest}) from $\beta_{left} = \beta^{\ast} - \Delta \beta
$ to $\beta_{right} =\beta^{\ast} + \Delta \beta $ are \footnote{The fit result around the critical point is stable when 0.005 $ \leq \Delta \beta \leq 0.006 $. In this range the agreement doesn't get worse than in eq. (\ref{peggio}) .}
\begin{equation}
1.797 \leq c_{1} \leq 1.968 \  , \quad c_{2}=-0.185 \ , \quad  0.106 \leq \alpha \leq 0.122  \label{predi}
\end{equation}%
the agreement appears to be very good when compared with the recent estimate
in \cite{BC02} in which the authors found $\alpha _{obs}\approx 0.114(6)$ so
that
\begin{equation}
\Delta \alpha \approx \frac{ \vert \alpha _{obs}-\alpha \vert }{\alpha _{obs}}\approx 
\frac{7}{100}
\label{peggio}
\end{equation}

\begin{figure}[!htp]
\caption{Bold points represent a sampling of Kallen-Lehman free energy $F_{3D}^{(\protect\zeta ^{\ast },\protect%
\nu ^{\ast },\protect\lambda ^{\ast })}(\protect\beta)$ while the continuous line corresponds to the critical
part of the expected free energy $F_{CRIT}$ versus $\protect\beta$.}
\label{fig-cp}
\begin{center}
\includegraphics{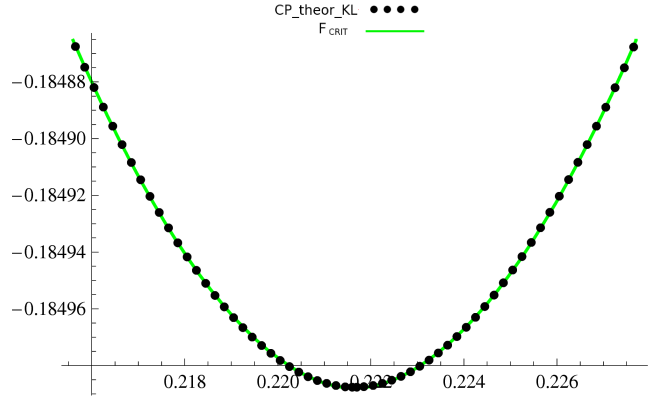}
\end{center}
\end{figure}

Further theoretical as well as experimental determinations of $\alpha $ can
be found in \cite{PV02}: the deviation $\Delta \alpha $ of our prediction (%
\ref{predi}) from observations appears to be less than 7\% in all the more
recent values. The following figure Fig.(\ref{fig-cp}) in which we draw both
the graph of the (non-analytic part of the) Regge free energy and of $%
F_{CRIT}$ confirm that in the parts of the graph of $F_{CRIT}$ in which the
dependence on $\alpha $ is important the agreement is satisfactory.

\section{Further possible improvements}

The first obvious improvement is to keep all the Regge exponents $\zeta _{3}$%
, $\zeta _{2}$, $\zeta _{1}$, and $\zeta _{0}$ without fixing \textit{a
priori} any of them (while here, for the sake of simplicity, we set both $%
\zeta _{3}$ and $\zeta _{1}$ to $1/2$, the 2D Ising values). As it has been
already stressed, it is also natural to explore different patterns of
duality breakings. One can keep in the parameters space the parameter $d_{0}$
without fixing it to 1 (the 2D Ising value). In the ansatz (\ref{parto32})
the terms corresponding to the 2D Ising model (\ref{t2D}) (which, in the
Kallen-Lehman view comes from the single particles discrete part of the
spectrum \cite{Ca07}) has a natural interpretation as a "boundary terms". In
other words, the 2D-like term with a single integral has its origin in the
boundary contributions in the same way as in the 2D case, the trivial
one-dimensional term has its origin in the boundary. It is conceivable that
the patterns of duality breakings in the bulk and at the boundary could be
different. This would lead to a Kallen-Lehman free energy in which the
function $k_{eff}(\beta )$ (see Eqs. (\ref{kaeff}), (\ref{co3}) and (\ref%
{co4})) appearing in the purely three dimensional term (the double integral)
has a $d_{0}$ different from the $k_{eff}(\beta )$ appearing in the two
dimensional term (the single integral in Eq. (\ref{t2D})). Two different $%
k_{eff}$ in the single and in the double integrals (each one with its own $%
d_{0}$) could describe a situation in which the pattern of duality breakings
in the bulk and in the boundary are different\footnote{%
Also worth to be investigated is the case in which $d_{0}$ has a smooth
dependence on the temperature in order to describe a pattern of duality
breaking which changes with $\beta $.}. We expect that more general duality
breakings could improve of two or three order of magnitude the (already
quite good) agreement at low temperatures. As a matter of fact, all these
natural ways to improve the numerical results (which add at least two
parameters to the computations) would need a big cluster of CPUs since
without a cluster the CPU's time needed would be too long.

\section{Conclusions and perspectives}

In this paper a Kallen-Lehman approach to 3D Ising model has been
investigated numerically in the realm of a minimal duality breaking. It has
been shown that one can fix three parameters of the model to get an
excellent agreement at high temperatures. \textit{With the same set of
parameters}, the agreement at low temperatures appears to be very good.
Remarkably enough, \textit{with the same set of parameters}, the predicted
critical exponent $\alpha $ has a relative deviation with respect to the
most recent determinations of the order of the 7\%. We believe that the
present results provide one with a strong evidence that the application of
the present methods to the study of the 3D Ising model is very promising. To
the best of our knowledge, there are no other analytical methods able to
give reliable informations at the same time at high temperatures, at low
temperatures and at the critical point. More general patterns of duality
breakings are worth to be investigated since they would further improve the
already very satisfactory agreement with MonteCarlo data.

\section*{Acknowledgements}

This work was supported by Fondecyt grant 3070055, 1071125, 1061291,
1051056, 1051064. The Centro de Estudios Cient\'{\i}ficos (CECS) is funded
by the Chilean Government through the Millennium Science Initiative and the
Centers of Excellence Base Financing Program of Conicyt. F.C. kindly acknowledeges Agenzia Spaziale Italiana for partial support. CECS is also
supported by a group of private companies which at present includes
Antofagasta Minerals, Arauco, Empresas CMPC, Indura, Naviera Ultragas and
Telef\'{o}nica del Sur. CIN is funded by Conicyt and the Gobierno Regional 
de Los R\'{\i}os. L. P. thanks PRIN SINTESI 2007 for financial support.

\end{document}